\documentclass[11pt,aps,amsmath,amssymb,nofootinbib,notitlepage,longbibliography]{revtex4-1}
\usepackage{amsmath,amssymb}
\usepackage{slashed}
\usepackage{graphicx}
\usepackage{bm}
\usepackage[top=1.0in,bottom=1.0in,left=1.0in,right=1.0in]{geometry}
\usepackage[colorlinks,linkcolor=blue,citecolor=blue]{hyperref}

\newcommand{\be}{\begin{equation}}
\newcommand{\ee}{\end{equation}}
\newcommand{\bea}{\begin{eqnarray}}
\newcommand{\eea}{\end{eqnarray}}
\newcommand{\ba}{\begin{eqnarray}}
\newcommand{\ea}{\end{eqnarray}}

\begin{document}

\title{Entanglement entropy and flow  in two dimensional QCD:\\
parton and string duality}

\author{Yizhuang Liu}
\email{yizhuang.liu@uj.edu.pl}
\affiliation{Institute of Theoretical Physics, Jagiellonian University, 30-348 Krak\'{o}w, Poland}

\author{Maciej A. Nowak}
\email{maciej.a.nowak@uj.edu.pl}
\affiliation{Institute
of Theoretical Physics and Mark Kac Complex Systems Research Center,
Jagiellonian University, 30-348 Krak\'{o}w, Poland}

\author{Ismail Zahed}
\email{ismail.zahed@stonybrook.edu}
\affiliation{Center for Nuclear Theory, Department of Physics and Astronomy, Stony Brook University, Stony Brook, New York 11794--3800, USA}



\begin{abstract}
We discuss quantum entanglement between fast and slow degrees of freedom,  in a two dimensional (2D)
large $N_c$ gauge theory with Dirac quarks, quantized on the light front.
Using the 't Hooft wave functions, we construct the reduced density matrix for an interval in the momentum fraction $x$-space,  and calculate its von Neumann entropy in terms of structure functions,  that are measured by DIS  on mesons (hadrons in general). We found that the entropy is bounded by an area law with logarithmic divergences, proportional to the rapidity of the meson. The evolution of the entanglement entropy with rapidity, is fixed  by the cumulative singlet  PDF,  and bounded from above by a Kolmogorov-Sinai
entropy of 1. At low-$x$, the entanglement exhibits an  asymptotic expansion,  similar to the forward meson-meson scattering amplitude
in the Regge limit. The  evolution of the entanglement entropy in parton-$x$ per unit rapidity, measures
the meson singlet PDF.  The re-summed
 entanglement entropy along the single meson Regge trajectory,
is string-like. We suggest that its extension
to multi-meson states, models DIS scattering on a large 2D $^\prime$nucleus$^\prime$. The
result, is a large rate of change of the entanglement entropy with rapidity,
that matches the current Bekenstein-Bremermann bound for maximum quantum information flow.
This mechanism may be at the origin of the  large entropy deposition and rapid thermalization,
reported in current heavy ion colliders, and may  extend to future electron-ion colliders.
\end{abstract}

\maketitle

\section{Introduction}
Quantum entanglement permeates most of our  quantum description of physical laws.
It follows from the fact that quantum states
are mostly superposition states, and two non-causally related measurements can be correlated,
as captured by the famed EPR paradox. A quantitative
measure of this correlation is given by the quantum entanglement entropy.  The entanglement entropy of quantum many body system and quantum field theory has been extensively explored in the literature~\cite{Srednicki:1993im,Calabrese:2004eu,Casini:2005rm,Hastings:2007iok,Calabrese:2009qy}. Less known perhaps, is the concept
of quantum entanglement flow, and its relation to
quantum information  flow and storage. A maximum flow is  expected in the most ideal
quantum systems, following from the bound in energy change  imposed by the  uncertainty principle~\cite{le1967proceedings,Bekenstein:1981zz}.

In hadron physics, quantum entanglement is inherent to a hadron undergoing large longitudinal boosts,
with its wavefunction described either by {\it wee partons}~\cite{Feynman:1969wa},
 or  {\it string bits}~\cite{Susskind:1993ki,Susskind:1993aa,Thorn:1994sw}. Entanglement
entropies are currently measured in diffractive $pp$ scattering at large $\sqrt{s}$
in current collider facilities~\cite{Stoffers:2012mn,Shuryak:2017phz,Liu:2018gae}, and will be measured in $ep$ scattering at low-x at future eIC facilities~\cite{Kharzeev:2017qzs,Stoffers:2012mn,Shuryak:2017phz,Liu:2018gae}, with better accuracy. Entanglement entropies in relation to hadronic processes, have also been discussed
in~\cite{Armesto:2019mna,Dvali:2021ooc}.

 In ultra-relativistic heavy-ion collisions, these inherently large entanglement entropies are at the
 origin of the prompt flow of {\it wee} entropy, likely at the boundary of our quantum laws. They may also explain,
  the almost instantaneous thermalization
 of the current strongly coupled plasma delivered initially at the RHIC facility, and later at the LHC facility~\cite{Stoffers:2012mn,Qian:2015boa,Shuryak:2017phz}.
 The duality between the low-x partons and the string bits~\cite{Karliner:1988hd,Qian:2014rda,Shuryak:2017phz}, explains why their entanglement
 provides for the most efficient mechanism for scrambling information,  matching only that produced
by gravitational black holes~\cite{Susskind:1993ki,Susskind:1993aa}.

In this work we discuss entanglement in longitudinal partonic momentum or Bjorken-x space, and also
in rapidity space or ${\rm ln}\frac 1x$ using 2-dimensional QCD. In the large number of colors limit, 2D QCD
is solvable with a dual partonic~\cite{tHooft:1974pnl} and string-like description~\cite{Bars:1976nk}. The purpose of this work is to elucidate the concept
of entanglement in single hadron states, or along a fixed Regge trajectory, as probed with DIS kinematics.
As an example of DIS scattering on a 2D nucleus, we will address the entanglement in a multi-meson state
(recall that all hadrons are similar on the light front),
and show how its growth rate in rapidity,  saturates the current bound on quantum information flow.

The outline of the paper is as follows: In section~\ref{LC2D} we briefly review the light cone formulation of 2D QCD
with Dirac quarks. In the large number of color limit, the two-body sector decouples and  solves the
't Hooft equation~\cite{tHooft:1974pnl}. In section~\ref{sec_EET},  we detail
the entangled density matrix in a single meson state,  with a single  parton-x cut, as probed by DIS scattering.
The evolution of the entanglement entropy with rapidity, is fixed by the cumulative PDF, obeys a Kolmogorov-Sinai bound of 1~\cite{Latora:1999vk} (and references therein),
and reduces to the
longitudinal meson structure function at low-x. The entanglement
entropy is  shown to be universal in the 2D scaling limit. In section~\ref{sec_2DIRAC} we recast 2D QCD as
a string on the light front in the 2-particle sector. We show that the entanglement of the string bits, follows
by resumming over the one meson Regge trajectory, thanks to duality. We suggest that the resummation over multi-meson
Regge trajectories may describe DIS scattering on a 2D {\it nucleus} on the  light front. The evolution of the ensuing
entanglement entropy with rapidity is extensive in the classical and longitudinal string entropy.
The rate of change matches the Bekenstein-Bremermann bound~\cite{le1967proceedings,Bekenstein:1981zz} for the maximum flow of
quantum information. Our conclusions are in section~\ref{sec_CON}. More details are given in the Appendices.

\section{Discrete Light-cone quantization of 2D QCD}~\label{LC2D}

To construct the reduced density matrix we first provide a review of the discrete light-cone quantization of the theory~\cite{Pauli:1985ps,Eller:1986nt,Eller:1989ek}. The system is put in a finite box in the light-front space $-\frac{L^-}{2}<x^-<\frac{L^-}{2}$. After choosing anti-periodic boundary condition, the momenta are labeled as
\begin{align}
k^+_p=\frac{\pi}{L^-}(2p+1) \ .
\end{align}
The good component $\psi_{+i}$ of the fermion field has the mode decomposition as
\begin{align}
\psi_{+i}(x^-)=\frac{1}{\sqrt{2L^-}}\sum_{p=0}^{N}\bigg( a_{i,p} e^{-i\frac{\pi (2p+1)}{L^-}x^-}+b_{i,p}^{\dagger}e^{i\frac{\pi (2p+1)}{L^-}x^-}\bigg) \ ,
\end{align}
which satisfies the anti-commutation relation
\begin{align}
[\psi_{+i}(x_1^-),\psi^{\dagger}_{+j}(x_2^-)]_{+}=\delta(x_1^--x_2^-)\delta_{ij} \ .
\end{align}
Here $i=1,..N_c$ is the color indices of the fermion field, which will be omitted below to avoid cluttering.  The total number of $N$ for a finite system with lattice cutoff $a$ is given by $N=[\frac{L^-}{2a}]-1$. Of all the $N$ independent frequencies, half are unfilled ($a_p$) and half are filled ($b_p$).  In terms of the above Light Front (LF) free field, the LF momentum $P^+$ and LF Hamiltonian are given by~\cite{Eller:1986nt,Eller:1989ek}
\begin{align}
P^+L^-=2\pi\sum_{p=0}^{N}(p+\frac{1}{2})\bigg(a_p^{\dagger}a_p+b_p^{\dagger}b_p\bigg) \ ,
\end{align}
and
\begin{align}
\frac{P^-}{L^-}\equiv H=\frac{M^2 }{2\pi}H_0+\frac{1}{L^-}V \ ,
\end{align}
with $H_0$ the mass contribution
\begin{align}
H_0=\sum_{p=0}^{N}\frac{a_p^{\dagger}a_p+b_p^{\dagger}b_p}{p+\frac{1}{2}} \ ,
\end{align}
Here $V$ consists of four-quark contributions which can be computed from the interaction term
\begin{align}
V=\frac{g_{1+1}^2}{2}\int_{-\frac{L^-}{2}}^{\frac{L^-}{2}} dx^- \psi_+^{\dagger}\psi_+ \frac{1}{(i\partial_-)^2} \psi_+^{\dagger}\psi_+ \ .
\end{align}
Expressed in terms of $a_p, b_p$, $H$ is independent of $L^-$. Using the explicit formula of $P^+$ above, it is clear that to describe a given hadron state with total momentum $P^+$, not all the modes are required. We only need those $p$ below
\begin{align}
p\le \frac{L^- P^+}{2\pi}-\frac{1}{2}\equiv \Lambda^--\frac{1}{2} \ .
\end{align}
Therefore,  $\Lambda^-$ provides a natural truncation of the Hilbert space. The momentum fractions are labeled by
\begin{align}
\frac{1}{2\Lambda^-}\le x_p=\frac{1}{2\Lambda^-}(2p+1)\le 1 \ .
\end{align}
Below, we use the label $x$ for all momenta. For a generic $\Lambda^-$, the states described above are purely discrete and breaks the Lorentz invariance. We expect that for $\Lambda^-\rightarrow \infty$, the spectrum of $H$ goes to zero as $\frac{M^2}{\Lambda^-}$,  and  the Lorentz invariant dispersion relation $P^+P^-=2M^2$ is restored.  In particular, the meson state can be constructed as
\begin{align}
|n\rangle=\frac{1}{\sqrt{\Lambda^-}}\sum_{0<p<\Lambda^-}\varphi_p a_p^{\dagger}b_{\Lambda^--p}^{\dagger}|0\rangle \ .
\end{align}
At large $N_c$, the above two-body state closes under the action of $P^-$. Requiring it to be an eigenstate of $P^-$ leads to the equation
\begin{align}
\label{PRETHOOFT}
(\Lambda^-)^2 m_R^2 \frac{\varphi_p}{(p+\frac{1}{2})(\Lambda^--p)}+\Lambda^-\frac{g_{1+1}^2N_c}{\pi}\sum_{l\ne p}\frac{\varphi_p-\varphi_l}{(p-l)^2}=M^2\varphi_p \ .
\end{align}
In the continuum limit $\Lambda^-\rightarrow \infty$, and with  the identification $x=\frac{p+\frac{1}{2}}{\Lambda^-}$ and $y=\frac{l+\frac{1}{2}}{\Lambda^-}$,
(\ref{PRETHOOFT}) reduces to the 't Hooft integral equation~\cite{tHooft:1974pnl} in the continuum
\begin{align}
\label{THOOFT}
\frac{m_R^2}{x\bar x}\varphi_n(x)+\frac{g_{1+1}^2N_c}{\pi}{\rm PV}\int_{0}^1 dy \frac{\varphi_n(x)-\varphi_n(y)}{(x-y)^2}=M_n^2\varphi(x) \ .
\end{align}
The gauge coupling is related to the string tension  $g_{1+1}^2N_c/2=\sigma_T$ (see below).
The renormalized quark mass is $m_R^2=m_Q^2-2\sigma_T/\pi$.
The ensuing spectrum is discrete,  with eigenvalues and eigenvectors labeled by  $M_n^2$ and $\varphi_n(x)$, respectively. They form a complete set of states in $L^2[0,1]$,
\begin{align}
\sum_n \varphi_n^{\dagger}(x)\varphi_n(x')=\delta(x-x') \ .
\end{align}
Their semi-classical and asymptotic behaviors are  briefly reviewed in Appendix~\ref{app_WKB}.

\section{Entanglement entropy in 2D QCD}~\label{sec_EET}

We now consider how different parts of a meson light front wave function as a bound quark-anti-quark state, are entangled
in the quark longitudinal momentum $k^+=xP^+$~\cite{Kharzeev:2017qzs,Shuryak:2017phz,Liu:2018gae,Kharzeev:2021nzh}. In particular, we will focus on the entanglement on a single asymmetric
cut in longitudinal momentum, by analogy with a DIS experiment where a single parton-x is singled out, say in the segment
$x_0\leq \frac 12$, including the low-x region.  We start by carefully reviewing the structure of the Hilbert space, and then
define the pertinent single cut entanglement entropy.

\subsection{Density matrix in longitudinal momentum}

 Since the color will be always traced out, here we simply omit the color factor. This will not modify our calculation of the entanglement entropy for two body states.  With this in mind, for each $x$ we have the quark and anti-quark operators $a_x,b_x$, and their corresponding 2D Fock space. The total un-constrained Hilbert-space is their tensor product
\begin{align}\label{eq:tensor}
{\cal H}=\bigotimes_{0<x<1} {\cal H}_x \otimes {\cal \bar H}_{x} \ ,
\end{align}
where
\begin{align}
{\cal H}_x={\rm Span}(|0\rangle_x ,a_x^{\dagger}|0\rangle_x) \ , {\cal \bar H}_x={\rm Span}(|\bar 0\rangle_x ,b_x^{\dagger}|\bar 0\rangle_x) \ .
\end{align}
The total dimension of the Hilbert space is then $2^{[\Lambda^--\frac{1}{2}]+1}\times 2^{[\Lambda^--\frac{1}{2}]+1} $, spanned by quark and antiquarks. In a confining theory, however, not all of the states in the above Hilbert space are physical. In the 2D QCD, it can be shown that in the large $N_c$ limit,  the physical spectrum consists of bound states formed by quark and anti-quarks, more precisely, the meson wave function reads
\begin{align}
|n\rangle=\frac{1}{\sqrt{\Lambda^-}}\sum_{0<x<1}\varphi_n(x)|x,\bar x\rangle \ ,
\end{align}
where the basis $|x,\bar x\rangle$ can be written in full tensorial form as
\begin{align}
|x,\bar x\rangle= a_x^{\dagger}|0\rangle_x\otimes b_{\bar x}^{\dagger}|\bar 0 \rangle_{\bar x} \otimes_{y\ne x}|0\rangle_y\otimes |\bar 0|\rangle_{\bar y} \ .
\end{align}
For finite $\Lambda^-$, $\varphi_n(x,\Lambda^-)$ satisfies a discrete version of the 't Hooft equation, but as $\Lambda^-\rightarrow \infty$
$\varphi_n(x,\Lambda^-)$ should converge to its continuum version given above. Below we will always use the continuum version of the wave function.  Unlike the free-quark and anti-quark states, the total dimension of the Hilbert space spanned by the 't Hooft wave functions is not ${\cal H}$, but only the two-quark states spanned by the set of bases $|x,\bar x\rangle$ defined above. Indeed, using the completeness equation of the 't Hooft equation one can show that
\begin{align}
\sum_n |n\rangle \langle n| =\sum_{0<x<1} |x,\bar x\rangle \langle x, \bar x| \ .
\end{align}
which is nothing but the projection operator into these quark-antiquark two body states. The total dimension of these states is only $\Lambda^-$, but not $4^{\Lambda^-}$.

Given the above meson state, one can construct its density matrix as
\begin{align}
\rho_n=\frac{1}{\Lambda^-}\sum_{x,x'}\varphi_n^{\dagger}(x')\varphi_n(x)|x,\bar x\rangle \langle x', \bar x'| \ .
\end{align}
Below we investigate its entanglement entropy with respect to the tensor product structure in Eq.~(\ref{eq:tensor}).

\subsection{Reduced density matrix}

The entanglement in longitudinal space is captured by the reduced matrix

\bea
\rho_{n}(x, x^\prime)={\rm tr}_A  \rho_n(x, x^\prime) \ .
\eea
where  $A$ denotes the part of the Hilbert space spanned by $a_x$, $b_x$,
 with $x$ lying in one or more sub-intervals of $[0,1]$.  How to choose $A$ depends on the probe experiment of interest. For instance, when  probing a hadron in a DIS experiment via  hard scattering, the virtual photon selects a quark or antiquark
with fixed parton-x, say $x_0<\frac 12$ in the range $\bar A=[0,x_0]$. The DIS event traces the hadron density matrix
over the remaining, and unobserved longitudinal momentum range $A=]x_0, 1]$. This is particularly clear, when probing
a hadron  using semi-inclusive DIS production of heavy mesons. In the large $N_c$ or planar approximation, the process is dominated by
Reggeon exchange, with the measured parton-x, kinematically limited to small $x_0\ll 1$. This reduction of the density matrix is asymmetric in parton-x.  A more symmetric but rather {\it academic} reduction,  is discussed in Appendix~\ref{app_AS}.

With this in mind, we now perform the partial trace in the tensor product in Eq.~(\ref{eq:tensor}), over all the ${\cal H}_x$ and ${\cal \bar H}_{x}$ where $x>x_0$
with $x_0<\frac 12$. To carry the partial trace, it is clear that for $x<x_0$ and $x'<x_0$, we are left with the quark contribution
\begin{align}
\frac{1}{\Lambda^-}\sum_{x<x_0}|\varphi_n(x)|^2a_x^{\dagger}|0\rangle_x \langle 0|_x a_x \bigotimes_{y<x_0,y'<x_0,y\ne x}|0\rangle_y|\bar 0\rangle_{y'}\langle \bar 0|_{y'}\langle 0|_y \ .
\end{align}
Similarly, for $x>1-x_0$ and $x'>1-x_0$, we have the anti-quark contribution
\begin{align}
\frac{1}{\Lambda^-}\sum_{x<x_0}|\varphi_n(\bar x)|^2b_x^{\dagger}|\bar 0\rangle_x \langle \bar 0|_x b_x \bigotimes_{y<x_0,y'<x_0,y'\ne x}|0\rangle_y|\bar 0\rangle_{y'}\langle \bar 0|_{y'}\langle 0|_y \ .
\end{align}
It is easy to see that in the cases where
$x<x_0,\,\,\,x'>1-x_0\qquad {\rm or}\qquad x>1-x_0,\,\,\,x'<x_0,$
 there are no partial traces that can be formed since in both of these two cases there will be one quark below $x_0$ and another quark above $x_0$. The case
 $x_0<x<1-x_0\qquad {\rm and} \qquad x_0<x'<1-x_0$
  should be considered, since in this case both the quark and antiquark are  above $x_0$,
   and should be traced out. This will leads to the contribution
\begin{align}
\frac{1}{\Lambda^-}\sum_{x_0<x<1-x_0}|\varphi_n(\bar x)|^2 \bigotimes_{y<x_0,y'<x_0}|0\rangle_y|\bar 0\rangle_{y'}\langle \bar 0|_{y'}\langle 0|_y \ .
\end{align}
The contribution is proportional to the vacuum contribution $|0\rangle \langle 0|$ for all the momentum modes below $x_0$ since they should not be traced over. Summing over the above, we found that for the two-body LF wave functions of a meson state, the reduced density matrix is diagonal and can be written schematically as
\begin{align}\label{eq:reduced}
\hat \rho_n({x_0})=\frac{1}{\Lambda^-}\sum_{x<x_0}\bigg[|\varphi_n(x)|^2|x\rangle_q\langle x|_q+|\varphi_n(\bar x)|^2|x\rangle_{\bar q}\langle x|_{\bar q}\bigg]+\frac{1}{\Lambda^-}\sum_{x_0<x<1-x_0}|\varphi_n( x)|^2|0\rangle\langle 0| \ .
\end{align}

The first contribution in (\ref{eq:reduced})
stems from the valence quark-antiquark pair in an n-meson state, and is expected. The second contribution
stems from the vacuum state (zero-modes) assumed normalizable, and is unexpected.
The trace of the reduced density matrix is $1$,  using the normalization condition of the wave function
\begin{align}
\sum_{0<x<1} \frac{\langle x|x\rangle}{\Lambda^-}|\varphi_n(x)|^2=1 \ .
\end{align}
with the light-like cutoff $\Lambda^-$

\bea
{\langle x|x\rangle}=2\pi\delta(0_x)=2\pi P^+\delta(0_{k^+})=\frac{P^+}{0_{k^+}}\equiv \Lambda^- \ .
\eea
From the light cone discretization of 2D QCD, we identify $0_{k^+}=1/2L^-$ as the lowest resolved longitudinal momentum, for a meson with total longitudinal momentum $P^+$.
 In the parton model,
$N=P^+/{0_{k^+}}$ counts the number of {\it wee} partons, with the larger the momentum, the larger $N$
(see also below).
We identify $\chi=\ln \Lambda^-$ with the rapidity, which is  fixed by DIS kinematics
as $\chi\sim \ln( Q^2/x)$ at low-$x$.

\subsection{Von Neumann entropy}

Given the reduced density matrix, the corresponding von Neumann  entanglement entropy is given by
\begin{align}
&S_n(x_0)=-{\rm tr} \hat \rho_n({x_0})\ln \hat \rho_n({x_0})= \ln\Lambda^-\int_{0}^{x_0} dx\bigg[|\varphi_n(x)|^2+|\varphi_n(\bar x)|^2\bigg]\nonumber \\
&-\int_{0}^{x_0} dx\bigg[|\varphi_n(x)|^2\ln |\varphi_n(x)|^2+|\varphi_n(\bar x)|^2\ln|\varphi_n(\bar x)|^2 \bigg]-\int_{x_0}^{1-x_0} dx |\varphi_n(x)|^2 \ln \int_{x_0}^{1-x_0} dx |\varphi_n(x)|^2 \ .
\end{align}
Since  the n-state quark and antiquark PDF for a meson is given by
\begin{align}
q_n(x)=\varphi_n^2(x) \ ,\qquad
\bar q_n(x)=\varphi^2_n(\bar x)\ ,
\end{align}
the entanglement entropy is specifically
\begin{align}
\label{SENTFULL}
S_n(x_0)=&\ln\Lambda^-\int_{0}^{x_0} dx \bigg[q_n(x)+\bar q_n(x)\bigg]-\int_{0}^{x_0} dx\bigg[q_n(x)\ln q_n(x)+\bar q_n(x)\ln \bar q_n(x) \bigg] \nonumber \\
&-\int_{x_0}^{\frac{1}{2}} dx \bigg[q_n(x)+\bar q_n(x)\bigg] \ln \int_{x_0}^{\frac{1}{2}} dx \bigg[q_n(x)+\bar q_n(x)\bigg] \ .
\end{align}
which is symmetric under the exchange of a  quark to an anti-quark.
Note that for $x_0=\frac{1}{2}$,  the result simplifies
\begin{align}
S_n\bigg(\frac{1}{2}\bigg)=\ln\Lambda^--\int_{0}^1 dx q_n(x)\ln q_n(x) \rightarrow \ln \Lambda^--(1-\ln 2) \ .
\end{align}
with the rightmost result following from the WKB approximation. We have checked that for other hadrons (nucleons, exotics),
the extensive part in (\ref{SENTFULL}) with the rapidity,  is also multiplied
by the cumulative probabilities of each parton in that state.
(\ref{SENTFULL}) is the first major result of this paper.
\\
\\
{\bf Area law and Kolmogorov-Sinai bound~\cite{Latora:1999vk}:}
\\
Since the entanglement entropy depends on the length of the interval $x_0\Lambda^-$ only through logs, it trivially satisfies an area law. Similarly to the spatial entanglement in 2D gapped system~\cite{Srednicki:1993im}, the entanglement contains a log-divergent term $\propto \ln \Lambda^-$  and a finite term.  However,
unlike the spatial entanglement entropy, the coefficient of the log-term depends also on the length of the interval.  The logarithmic dependence leads to an evolution in rapidity, and is bounded from above as
\begin{align}
\label{BOUND}
\frac{d S_n(x_0)}{d\chi}=\int_{0}^{x_0} dx \bigg[q_n(x)+\bar q_n(x)\bigg]\equiv C(x_0) \leq 1 \ ,
\end{align}
In a way, the analogue of the {central charge}, is played by the {\it cumulative parton probability}  $C(x_0)$,
with $C(\frac 12)=1$ saturating the bound.

If we identify  the logarithmic dependence on $P^+$ as an evolution in rapidity, then  (\ref{BOUND}) can be viewed as the Kolmogorov-Sinai bound for the entanglement entropy for an n-meson in 2-dimensional QCD, and identify the Kolmogorov-Sinai entropy
$S_{KS}=1$ (sum of the positive Lyapunov exponents).

The bound (\ref{BOUND}) can  be understood in the following way.  For the 't Hooft wave functions, the reduced density matrix contains only one-body and zero body (vacuum) terms, therefore its Schmidt decomposition allows at most $2x_0\Lambda^-$ terms,  which implies an upper-bound $S_n \le \ln \Lambda^-+\ln 2x_0$. However, our result shows  that this is an  over-estimate. For the two-body wave function, it is the finite probability of the  zero-mode contribution (vacuum state), that  reduces  the over-estimation.  For three- and higher-body wave functions, we show in Appendix \ref{App_generic},
that the naive  upper bound $\propto (k-1)\ln \Lambda^-$, for a generic state where $k$ is the maximal number of
partons, is also an over-estimate.
\\
\\
{\bf Structure function:}
\\
At  low-x,  (\ref{BOUND}) is the n-meson $F^n_2$ structure function

\begin{align}
\label{BOUNDF}
\frac{d S_n(x_0\sim 0)}{d\chi}\sim x_0 \big(q_n(x_0)+\bar q_n(x_0)\big) =F^n_2(x_0\sim 0) \ ,
\end{align}
in agreement with the  analysis in higher dimensions~\cite{Stoffers:2012mn,Kharzeev:2017qzs,Shuryak:2017phz,Liu:2018gae}. In 2D (\ref{BOUNDF})  measures  the low-x partons in the  n-meson state

\begin{align}
\label{BOUNDX}
\frac{d S_n(x_0\sim 0)}{d\chi}\sim 2C_n^2\frac{x_0^{2\beta+1}}{2\beta+1} \ ,
\end{align}
where we used that at the edges $x=0$ and $x=1$. The  't Hooft wave function has an asymptotic expansion,  in terms of the dynamically generated coefficient $\beta$ as
\begin{align}
\label{WFS}
\varphi_n(x)=C_n x^{\beta}, \qquad \pi \beta \cot \pi \beta=-\frac{\pi m_Q^2}{2\sigma_T}+1 \ .
\end{align}
A more refined analysis detailed in Appendix~\ref{app_EXP},  gives
\bea
\label{SNX0}
S_n(x_0)=2C_n^2\frac{x_0^{2\beta+1}}{2\beta+1}
\bigg(\ln (e\Lambda^-)+2\beta\frac{1+(2\beta+1)\ln \frac{1}{x_0}}{(2\beta+1)}+{\cal O}(x_0^2)\bigg) \ .
\eea
for $\beta>0$.
The result is  consistent with (\ref{BOUNDF}), if we note that the  second contribution in (\ref{SNX0}) is suppressed in the chiral limit, i.e.  $\beta\sim m_Q/\sqrt{\sigma_T}$.
In passing, we also note  the non-commutativity of the chiral limit with the low-x limit in 2D QCD.

For theories in which there are non-trivial logarithmiths running in rapidity, for example  4-dimensional QCD, (\ref{BOUNDF})
measures the growth of low-x partons carried by the quark sea.
This is  consistent with the forward meson-meson (elastic n-n$\rightarrow$n-n)
scattering amplitude in the Regge limit in 2D~\cite{Brower:1978wm}
\bea
\sigma_n(s)\sim \frac 1s {\rm Im} A_n(s,0)\sim s^{-(2\beta+1)}\sim F^n_2(x_0\sim 0) \ ,
\eea
with a negative Reggeon intercept $\alpha_{\mathbb R}=-2\beta$ (In 4D the forward limit is dominated by the Pomeron with positive intercept $\alpha_{\mathbb P}>0$,
with the stringy relation $\alpha_{\mathbb R}+1=\alpha_{\mathbb P}$).
The forward elastic cross section,
is a measure of the n-meson $F^n_2$ structure function.
It is also consistent with the elastic 2D n-meson form factor $F_n(-q^2)\sim 1/(-q^2)^{\beta+1}$, both of which are dominated by the t-channel
{\it single} Reggeon   exchange, which amounts  to a full quantum
open string exchange after re-summation, as we show below.
\\
\\
{\bf Valence PDF:}
\\
The longitudinal evolution of the entanglement entropy (\ref{SENTFULL}) with parton-$x$, is highly non-linear
\bea
\frac{dS_n(x_0)}{dx_0}=&&-\big(q_n(x_0)\ln q_n(x_0)+\bar q_n(x_0)\ln \bar q_n(x_0) \big) \nonumber\\
&&+ \big(q_n(x_0)+\bar q_n(x_0)\big)
\ln \bigg(\frac{ \Lambda^-}{e\int_{x_0}^{\frac{1}{2}} dx \big(q_n(x)+\bar q_n(x)\big)}\bigg) \ ,
\eea
with most of the non-linearity arising from the entanglement with the vacuum contribution
in (\ref{eq:reduced}).
For large rapidities $\chi$, the longitudinal growth per unit rapidity is linear, and is a direct
measure of the $n$-meson valence PDF
\bea
\frac{d^2S_n(x_0)}{d\chi dx_0}=q_n(x_0)+\bar q_n(x_0) \ .
\eea
We expect a similar relation to hold  for more general wave functions, e.g.  baryons and exotics.
\\
\\
{\bf Scaling  limit:}
\\
Another interesting limit is the so-called  scaling limit, which consists on zooming on the large-n meson states
to exhibit the scale invariance of 2D QCD~\cite{Brower:1978wm,Katz:2007br}. More specifically, consider the limit
 $\mu^2_n=M_n^2/m_0^2\rightarrow \infty$ with fixed ratio  $\xi=x\mu_n^2$, where $m_0^2=2\sigma_T/\pi$.
 In this limit, the wave function
approaches a universal function $\phi(\xi)$:
\begin{align}
\varphi_n\bigg(\frac{\xi}{\mu_n^2}\bigg)\rightarrow \phi(\xi) \ .
\end{align}
In this case, if  we set  $x_0=\frac{\xi_0}{\mu_n^2}$, then  the cumulative parton distribution
\begin{align}
\int_{0}^{x_0} dx (q_n+\bar q_n)=\frac{2}{\mu_n^2}\int_{0}^{\xi_0} d\xi \phi^2(\xi)\ ,
\end{align}
and
\begin{align}
\int_{0}^{x_0} dx q_n\ln q_n=\int_{0}^{x_0} dx \bar q_n\ln \bar q_n=\frac{1}{\mu_n^2}\int_{0}^{\xi_0} d\xi \phi^2(\xi)\ln \phi^2(\xi) \ .
\end{align}
The leading ${\cal O}({1}/{\mu_n^2})$ entanglement entropy in this case is therefore purely expressed in terms of the universal function
\begin{align}
\mu_n^2 S_n\bigg(\frac{\xi_0}{\mu_n^2}\bigg)\rightarrow  2\ln e\Lambda^- \int_{0}^{\xi_0} d\xi \phi^2(\xi)-2\int_{0}^{\xi_0} d\xi \phi^2(\xi)\ln \phi^2(\xi) \ .
\end{align}
For large $\xi_0$, the first term diverge linearly in $\xi_0$, while the second term diverges logarithmically.
\\
\\
{\bf Pseudo-Goldstone  state:}
\\
In  the limit $m_Q^2/2\sigma_T\ll 1$, 2D QCD admits a massless pseudo-Goldstone mode with a light front
wavefunction $\varphi_0(x)=\theta (x\bar x)$ (modulo the end points).
This is a Berezinski-Kosterlitz-Thouless state. In 4-dimensional QCD, the pion is a true Goldstone mode,
and massless even for a fixed and large constituent mass $m_Q$, yet the pion longitudinal wave-function is also totally delocalized
in x-Bjorken with $\varphi_\pi(x)\approx \theta(x\bar x)$ in the chiral limit, and for point-like interactions~\cite{Kock:2020frx} (and references therein).

With this in mind,  the density matrix for the pseudo-Goldstone state reads
\bea
\label{RHOPI}
\hat \rho_{\pi}=\frac 1{\Lambda^-}\sum_{n=1}^\infty\sum_{x,x^\prime}\theta(x\bar x)\theta(x^\prime\bar{x}^\prime)
| x,\bar x\rangle \langle x^\prime, \bar x^\prime| \ .
\eea
modulo the end-points.
When traced over the interval $\bar x_0=1-x_0$, the
entanglement entropy  is

\bea
\label{SPION}
S_\pi(x_0)=2x_0 {\rm ln}\Lambda^--(\bar x_0-x_0){\ln}(\bar x_0-x_0) \ ,
\eea
which is considerably simpler than (\ref{SENTFULL}).
The change in rapidity of the pion entanglement entropy$$\frac{d S_\pi(x_0)}{d\chi}=2x_0$$
This result is similar to the one we derive below for the entanglement entropy summed over the full Regge trajectory.
This is perhaps the signature  of the collective nature of the pseudo-Goldstone mode,  on the light front.
We note  that in the massless Schwinger model, the light front  wave function of the ``meson'' state
with mass $m^2={g^2}/{\pi}$, is
\begin{align}
|\gamma \rangle=\frac{1}{\sqrt{\Lambda^-}}\sum_{0<x<1} |x,\bar x \rangle \ .
\end{align}
with the same entanglement entropy (\ref{SPION}) as in the pseudo-Goldstone state.


%

\section{2D QCD as a string  on the light front}~\label{sec_2DIRAC}

Two-dimensional QCD is non-conformal, but solvable in the large number of  colors limit~\cite{tHooft:1974pnl}, as we discussed
using the discretized light front quantization earlier.
Remarkably, the solution in this limit is identical to that following from a 2-dimensional relativistic string with
massive end-points~\cite{Bars:1976nk}.  To show this, we recall that the 2D light front Hamiltonian (squared mass) for a string with massive ends is~\cite{Bars:1976nk}

\bea
\label{HLF}
H_{LF}=\frac {m_Q^2}{x\bar x}+2P^+\sigma_T\big|r^-\big|\rightarrow \frac {m_Q^2}{x\bar x}+2\sigma_T\bigg|\frac{id}{dx}\bigg| \ ,
\eea
with $0\leq x=k^+/P^+\leq 1$ the momentum fraction of the quark
($\bar x=1-x$ is that of the anti-quark) in a meson with longitudinal momentum
$P^+$.  The relative light-front distance $r^-\rightarrow id/dk^+$
is conjugate to $k^+$. The string tension is $\sigma_T$. The eigenstates of (\ref{HLF}) solve

\bea
\label{HLFX}
H_{LF}\varphi_n(x)=\bigg(\frac {m_Q^2}{x\bar x}+2\sigma_T\bigg|\frac{id}{dx}\bigg|
\bigg)\varphi_n(x)=M_n^2\varphi_n(x) \ ,
\eea
with squared radial meson masses as eigenvalues.
The confining potential  in the Bjorken-x representation is given by the Fourier transform

\bea
\label{CONF}
\langle x|P^+|r^-||y\rangle =\int_{-\infty}^{+\infty}\frac {dq}{2\pi}e^{iq(x-y)}|q|
\rightarrow {\rm PV}\frac {-1}{\pi(x-y)^2}+\frac {-1}{\pi x\bar x} \ ,
\eea
with the principal value prescription. Using (\ref{CONF}) in (\ref{HLFX}) yields 't Hooft equation (\ref{THOOFT})
with  the gauge coupling identified through $\sigma_T=g_{1+1}^2N_c/2$.
A brief semi-classical analysis of the string states is given in Appendix~\ref{app_WKB}.
In sum, we can regard the even and odd  solutions of the 't Hooft equation, as the even and odd standing waves of
a meson  as a string, flying on the light front with either Dirichlet or Neumann boundary conditions modulo the small mass corrections
at the edges.

\subsection{Stringy entanglement: Resummed Regge trajectory}~\label{sec_ONE}

In the eikonalized approximation, dipole-dipole (open string) scattering in 2D QCD, sums over all n-meson (Reggeons)
 exchanges in the t-channel.
This re-summed exchange is string-like.  To describe it, we need to resum over the full meson Regge trajectory in 2D QCD.
However, this is not needed as we now show.

Indeed, the full
density matrix of the string $\hat\rho_{\rm string}$, can be reconstructed from the n-meson density matrix $\rho_n$,
by noting that each of the  meson state on the Regge trajectory, maps onto a stationary state of the open string with massive end-points.
The orthonormality and completeness of these states, imply that  the full string density matrix is diagonal in-$n$,

\bea
\label{RHOSTRING}
\hat \rho_{\rm string}=\frac 1{\Lambda^-}\sum_{n=1}^\infty\sum_{x,x'}\varphi_n^{\dagger}(x')\varphi_n(x)| x,\bar x\rangle \langle x^\prime, \bar x^\prime| \ .
\eea
Using the completeness relation
\begin{align}
\sum_n \varphi_n^{\dagger}(x)\varphi_n(x')=\delta(x-x') \ ,
\end{align}
(\ref{RHOSTRING}) is  the projection operator onto the two-body states
\begin{align}
\hat \rho_{\rm string}=\frac{1}{\Lambda^-}\sum_{0<x<1} |x,\bar x\rangle \langle x,\bar x| \ .
\end{align}
The reduced density matrix,  following by tracing over the segment $\bar x_0=1-x_0$, yields the entanglement entropy
\begin{align}
S(x_0)=2x_0\ln \Lambda^--(\bar x_0-x_0)\ln (\bar x_0-x_0) \ .
\end{align}
which is independent of the mass at the end-points of the string.
It is surprisingly similar to  (\ref{SPION}) for the  pseudo-Goldstone mode, even though the string density matrix (\ref{RHOSTRING}) is diagonal
in longitudinal space,  while the one associated to the pseudo-Goldstone mode (\ref{RHOPI}) is off-diagonal.

\subsection{Stringy entanglement: Multi-meson state}~\label{sec_MANY}

The above density matrix takes into account only single meson states. As we argued earlier, this entangled density matrix
captures a DIS measurement of the quark distribution in a meson state, in the interval of length $x_0$ in parton-x. Suppose that we
want to use a DIS measurement of the quark distribution for the same $x_0$ interval,  in a state composed of many identical
hadrons flying  on the light front (a 2D nucleus, or a 4D nucleus reduced to its longitudinal components). For that, we extend our analysis to multi-meson states, with the corresponding
 Fock-space spanned by all the mesons. Using the completeness relation, it is clear that the corresponding density matrix is now
 given by

\begin{align}
\rho=\frac{1}{\rm Dim}\sum_{k}\rho_k \ ,
\end{align}
where
\begin{align}
 \rho_k=\frac{1}{\rm Dim}\sum_{0<x_1<x_2..<x_k<1}|x_1,x_2,...x_k\rangle \langle x_1,...x_k| \ ,
\end{align}
are spanned by all $k$-particle tensor product of the fundamental basis, under the constraint that the same $|x,\bar x\rangle$ appears at most $N_c$ times. For large $N_c$, each can appear infinitely many times.
\\
\\
{\bf $N_c=1$ case:}
\\
To help understand the book-keeping for general $N_c$,
let us first consider the case with $N_c=1$, with no
2 mesons allowed to occupy the same longitudinal phase space region.  In this case ${\rm Dim}=2^{\Lambda^-}$.
After tracing over the segment $(1-x_0)$, the reduced density matrix for this case is
\begin{align}
\frac{1}{2^{\Lambda^-}}\sum_{k-(\Lambda^--N_1)<i<{\rm min}(k,N_1)}C_{N_1}^i|x_1,x_2,...x_{k-i}\rangle \langle x_1......x_{k-i}| \ ,
\end{align}
where $N_1=(1-x_0)\Lambda^-$.  After summing over all $k$ with the help of the binomial theorem, and
replacing $k-i$ by $\tilde k$,  the result is
\begin{align}
\rho(x_0)&=\frac{1}{2^{\Lambda^-}}\sum_{0\le \tilde k \le x_0\Lambda^-}\sum_{k=\tilde k}^{N_1+\tilde k}C_{N_1}^{k-\tilde k}\sum_{0\le x_1 <x_2....<x_{\tilde k}<x_0}|x_1,..x_{\tilde k}\rangle \langle x_1,....x_{\tilde k}| \ , \\
&\equiv \frac{1}{2^{x_0\Lambda^-}}\sum_{0\le \tilde k \le x_0\Lambda^-}\sum_{0\le x_1 <x_2....<x_{\tilde k}<x_0}|x_1,..x_{\tilde k}\rangle \langle x_1,....x_{\tilde k}| \ .
\end{align}
This is simply the projection operator onto the subspace with $x_0\Lambda^-$ digits, corresponding to the part of the Hilbert space kept. The dimension of the space is ${\rm Dim}(x_0)=2^{x_0\Lambda^-}$, and the corresponding entanglement entropy is now
\begin{align}
S_E=\ln {\rm Dim}(x_0)= \ln 2 \times x_0 \Lambda^-.
\end{align}
This is  the {\it maximal entropy},  following from the reduction of  any density matrix to the small-$x$ interval.
\\
\\
{\bf General $N_c$ case:}
\\
For general $N_c$, and after tracing over the $(1-x_0)$, we  clearly get again the projection operator onto the subspace spanned by all the $|x_1,...x_k\rangle$, with the constraint that $x_k\le x_0$ and that each $x_i$ appears at most $N_c$ times, due to the fermionic character of the underlying quark constituents in any of the colorless meson. The dimension of this Hilbert space is simply $(N_c+1)^{x_0\Lambda^-}$, hence
\begin{align}
S_E=\ln (N_c+1) \times x_0 \Lambda^- \ .
\end{align}

The rate of change with rapidity of the string entanglement entropy $S_E(x_0)$,
the sum total of  all entanglements along each of the exchanged  Regge trajectories for fixed $x_0\leq \frac 12$,  is extensive in $\Lambda^-$
\begin{align}
\label{SUMX}
\frac{dS_E(x_0)}{d\chi}=\ln (N_c+1) x_0 \Lambda^- \ .
\end{align}
In the low-x regime, dominated by the vacuum zero-modes on the light front, (\ref{SUMX})  simplifies to
\bea
\label{LOWX0}
\frac{dS_E(x_0\sim 0)}{d\chi}= \ln (N_c+1) \frac 12 e^{-\chi}e^{\chi}=\frac{1}{2}\ln (N_c+1) \ ,
\eea
using the DIS  identification $x_0=\frac{1}{2}e^{-\chi}$.
\\
\\
{\bf Kolmogorov-Sinai bound~\cite{Latora:1999vk}:}
\\
The rate of increase of $S_E(x_0\sim 0)$  with the rapidity $\chi$, saturates the Kolmogorov-Sinai bound at low-x,
with  $S_{KS}=\frac 12 \ln (N_c+1)$. The
 longitudinal quantum entanglement, for the re-summed mesons (Reggeon) as open strings  in 2D, is to be compared to
 the transverse quantum entanglement of $\frac {D_\perp}6$ for the re-summed glueballs (Pomeron) as a closed string
 exchange  in  $2+D_\perp$ dimensions~\cite{Stoffers:2012mn,Shuryak:2017phz,Liu:2018gae}.
At low-x, the entanglement
is fixed by the $D_\perp$ transverse quantum vibrations of the string light-like  (analogue of Luscher term space-like).
\\
\\
{\bf Classical string entropy:}
\\
Away from low-x, the change in $S_E(x_0)$ is extensive in
the invariant cut-off $\Lambda^-$, e.g.
$$\frac{dS_E(x_0)}{d\chi}=\ln (N_c+1) x_0\Lambda^-$$
This scaling is commensurate with  the growth of the string entropy $S_S$ under large boosts. Indeed,
a  {\it free} string as a chain undergoing random walks in
1D, generates  $N_S=2^{L/l_S}$ states (for a free string back-tracking is allowed). The corresponding
string entropy $S_S={\ln N_S}={\ln 2}\,L/l_S$. Under large longitudinal boosts $P^+$, the
longitudinal length of the string {\it expands}  (recall that the string bits are
considered {\it wee}~\cite{Karliner:1988hd,Bergman:1997ki},  they carry low momentum, and are oblivious to large boosts). As a result, ${L}/{l_S}= {P^+}/{0_{k^+}}=x_0\Lambda^-$ counts the number
of string bits or wee partons, and the string entropy is
$S_S=\ln2\,  x_0\Lambda^-$, which is seen to scale similarly to  (\ref{SUMX}), in particular
\bea
\label{MANYX}
\frac{dS_E(x_0)}{d\chi}=\frac {\ln (N_c+1)}{\ln 2} S_S  \ .
\eea
This large and quantum {\it wee} entropy stored in the longitudinal evolution in rapidity of open strings (Reggeons),
when released in a collision, may contribute to the fast scrambling of information in hadronic collisions at
ultra-relativistic energies. Perhaps more so, then the quantum  {\it wee} entropy released  from the evolution in rapidity of closed
strings (Pomerons)~\cite{Stoffers:2012mn}, provided that $x_0$  is not asymptotically small as in (\ref{LOWX0}).
We note that the string bits interactions may hamper the back-tracking, and somehow reduce the entanglement rate in (\ref{MANYX}).
\\
\\
{\bf Bekenstein-Bremermann bound~\cite{le1967proceedings,Bekenstein:1981zz}:}
\\
Quantum information theory sets a bound on the maximum rate of flow of information $I$ in physical systems, as first noted
by Bremermann for single channel systems, based on an argument using Shannon entropy and the quantum uncertainty principle~\cite{le1967proceedings}. The bound
was revisited  by Bekenstein on general grounds, using the maximum entropy  storage in a black-hole and causality~\cite{Bekenstein:1981zz}
\bea
\label{BEK}
\frac {dS_{\rm max}}{dt}\leq {2\pi E}\rightarrow {2\pi TS} \ .
\eea
The rightmost equality follows from  the second law.
(Here information $I$ is interpreted as entropy in  bits units or  $I/S= \ln_2e$).
If we recall that the rapidity $\chi$ relates to the Gribov time $t_\chi=\sqrt{\alpha^\prime}\chi$
with $\alpha^\prime=l_S^2$ the open string Regge slope~\cite{Stoffers:2012zw,Liu:2018gae}, then a comparison of (\ref{MANYX})
with (\ref{BEK}) shows that for $N_c=1$,
the Bekenstein-Bremermann bound is saturated, with $T=T_H=1/(2\pi l_S)$ the
Hagedorn temperature (equivalently, the temperature at the Rindler horizon of a black-hole).
Remarkably, for the multi-meson state result with $N_c>1$ in (\ref{MANYX}), the bound is still maintained,
provided that the temperature exceeds (logarithmically) the Hagedorn temperature.

\section{Conclusions}~\label{sec_CON}

In the large number of colors, the 2-particle sector of 2D QCD on the light front decouples. The eigen-modes
in this sector, have a dual description in terms of partons or string modes. We have shown that in the
partonic language, the entanglement in longitudinal momentum is captured by an exact reduced density
matrix, that is a tensor product of both the valence and vacuum states. The entanglement entropy for a
single meson with a single cut in parton-x, as probed by DIS kinematics, is  a non-linear function of
the meson PDF.

For fixed parton-x, the evolution in rapidity of the single meson entanglement entropy, is the {\it cumulative} quark single PDF.  It is bound
by a  Kolmogorov-Sinai entropy of 1. At low parton-x, it reduces to the longitudinal structure function, as measured in DIS scattering.
It is in agreement with the Regge behavior of the pertinent meson-meson scattering in 2D QCD. Alternatively, for fixed rapidity,
the evolution in parton-x is shown to probe directly the meson singlet PDF.

The sum total of the entanglement entropies for a fixed Regge trajectory, is string-like and extensive with the rapidity, as noted
in 4D. We have suggested that DIS scattering on a {\it nucleus} in 2D,  can be modeled by DIS scattering on a multi-hadron state
composed of 2D mesons, modulo Fermi statistics (amusingly shared by mesons through longitudinal space exclusion for $N_c=1$).
The evolution in rapidity of the ensuing entanglement entropy, is found to be extensive in the longitudinal string entropy in 2D.
The rate of change of this entropy matches the maximum rate of quantum information flow, as given by the Bekenstein-Bremermann
bound.

A highly boosted multi-meson  state in 2D (a sort of 2D {\it nucleus} as all hadrons are similar on the light front), exhibits a growth rate in its {\it wee} parton
entanglement entropy, that is only matched by the largest information rate flow allowed by the quantum laws of physics,
a fit only exhibited by gravitational black holes.
Remarkably, this flow exhibits an energy cost which is fixed by the Hagedorn temperature of the underlying longitudinal
string.

The highly  entangled {\it wee} partons in a boosted string as a {\it mock nucleus}, carry an entanglement entropy
that is commensurate with  the classical string entropy $S_S$.
Their prompt release by smashing,  in current colliders at large rapidities $\chi=\ln s$,  may explain why
a large quantum {entanglement} entropy of about $\chi S_S$ is promptly released, over a short time scale $1/l_S$,  and
at  temperatures in (slight) excess of the Hagedorn temperature $T_H=1/(2\pi l_S)$.

We will elaborate further on some of these issues and their extension to 4D next.

\vskip 1cm
{\bf Acknowledgements}

This work was supported by the U.S. Department of Energy under Contract No.
DE-FG-88ER40388, and  by the Priority Research Area SciMat under the program
Excellence Initiative – Research University at the Jagiellonian University in Kraków.

\appendix

\section{WKB analysis of the string states}~\label{app_WKB}

In this Appendix, we qualitatively review the semi-classical solutions to 2D QCD, using
the dual string form. In particular, the masses are given by the WKB quantization condition

\bea
\int_{x_-}^{x_+}dx \bigg(M_n^2-\frac{m_Q^2}{x\bar x}\bigg)=
M_n^2-m_Q^2\,{\rm ln}\bigg(\frac{x_+\bar{x}_-}{x_-\bar{x}_+}\bigg)=2\pi\sigma_T n \ ,
\eea
with the turning points
\be
x_\pm=\frac 12\bigg(1\pm \bigg(1-\frac{4m^2_Q}{M_n^2}\bigg)^{\frac 12}\bigg) \ ,
\ee
and with  $M_n\geq 2m_Q$. The mass gap vanishes  for $m_Q\rightarrow 0$ with  a radial Regge trajectory $M_n^2= n/\alpha^\prime$,
and $\alpha^\prime=1/2\pi\sigma_T$ the slope of the open bosonic string.

A simple understanding of the light front wavefunctions, can be obtained directly from (\ref{HLFX}) by noting
that for $m_Q^2/2\sigma_T\gg 1$, the mass contribution acts as a confining potential at the end-points $x=0,1$,
with $\varphi_n(x)$ standing waves solutions to

\bea
\bigg|\frac{id}{dx}\bigg|
\varphi_n(x)\approx \frac{M_n^2}{2\sigma_T}\varphi_n(x) \ ,
\eea
with Dirichlet boundary conditions. The normalized solutions are
 $\varphi_n(x)\approx \sqrt 2 \,{\rm sin}((n+1)\pi x)$.   
  A simple estimate
 of the mass correction for large $n$ follows from first order perturbation
 theory $M_n^2\approx n/\alpha^\prime +2m_Q^2\,{\rm ln}\,n$.
In  the opposite limit of $m_Q^2/2\sigma_T\ll1 $, the confining potential can be ignored to
 first approximation, in which case the standing waves follow from Neumann boundary conditions,
 with $\varphi_n(x)\approx \sqrt{2}{\rm cos}(n\pi x)$, with an identical reggeized semi-classical spectrum.
 The effects of the mass is to cause a rapid distortion of the light front wavefunction in a narrow region
 of $x$ near the end-points (see below).

\section{Symmetric interval}~\label{app_AS}

The reduced density matrix in parton-x, was defined by tracing over the length  $\bar x_0=1-x_0$ for fixed $x_0\leq \frac 12$,
as motivated by a DIS measurement.
This reduction is asymmetric with respect to the quark-antiquark content of the light front meson wavefunction. A more {\it symmetric}
but {\it academic} reduction,  is to trace over the symmetric length $x_0<x<\bar x_0$.  The reduced density matrix  is then

\begin{align}
\hat \rho_{S}(n)=\int_{x_0}^{\bar x_0}q_n(x) |0\rangle_{S} \langle 0|_{S}+ |\tilde \Phi\rangle \langle \tilde \Phi| \ ,
\end{align}
where one has
\begin{align}
|\tilde \Phi\rangle = \frac{1}{\sqrt{\Lambda^-}}\bigg(\sum_{0<x<x_0}+\sum_{\bar x_0<x<1}\bigg) \varphi_n(x)|x,\bar x\rangle  \ .
\end{align}
The above density matrix represents a binomial distribution,  with the  independent pair of  eigenvalues $(p_n(x_0),1-p_n(x_0))$ where
\begin{align}
p_n(x_0)=\int_{0}^{x_0} dx \bigg(q_n(x)+\bar q_n(x)\bigg) \ .
\end{align}
The corresponding  entanglement entropy is therefore
\begin{align}
S_{S}(n,x_0)=-p_n(x_0)\ln p_n(x_0)-(1-p_n(x_0))\ln (1-p_n(x_0)) \ .
\end{align}
and is independent of $\Lambda^-$. As $x_0 \rightarrow 0$, one has
\begin{align}
p_n(x_0)\rightarrow \frac{2C_n^2x_0^{2\beta+1}}{2\beta+1} \ ,
\end{align}
thus
\begin{align}
S_{S}(n,x_0)=2C_n^2x_0^{2\beta+1}\ln \frac{1}{x_0}-\frac{2C_n^2x_0^{2\beta+1}}{2\beta+1}\ln \frac{2C_n^2}{e(2\beta+1)}+{\cal O}(x_0^{4\beta+2}) \ .
\end{align}
The leading contribution is also  proportional to $x_0^{2\beta+1}\ln \frac{1}{x_0}$.

\section{General interpolating interval}
In this Appendix,  we  trace over an {\it asymmetric} interval centered around $\frac 12$, that interpolates between the symmetric and asymmetric
eduction discussed above. In this case,
the reduced density matrix traced over $[x_0,\bar{x}_0+\delta]$ with $0<\delta<x_0$, is now
\begin{align}
&\hat \rho=\frac{1}{\Lambda^-}\sum_{x_0<x<1-x_0}|\varphi_n(x)|^2 |0\rangle_{[x_0,1-x_0]}\langle 0|_{[x_0,1-x_0]}+\frac{1}{\Lambda^-}\sum_{x_0-\delta<x<x_0}|\varphi_n(x)|^2|x\rangle \langle x| \nonumber \\ +&\frac{1}{\Lambda^-}\sum_{1-x_0<x<1-x_0+\delta}|\varphi_n(x)|^2|\bar x\rangle \langle \bar x| +|\tilde \Phi\rangle \langle \tilde \Phi| \ ,
\end{align}
where the state $|\tilde \Phi \rangle$ reads
\begin{align}
|\tilde \Phi\rangle =\frac{1}{\Lambda^-}\bigg(\sum_{0<x<x_0-\delta}+\sum_{1-x_0+\delta<x<1}\bigg)\varphi_n(x)|x,\bar x\rangle \ .
\end{align}
The entanglement entropy is therefore given by
\begin{align}
&S_n(x_0,\delta)=\ln \Lambda^- \int_{x_0-\delta}^{x_0}dx \bigg(q_n(x)+\bar q_n(x)\bigg)-\int_{x_0-\delta}^{x_0} dx\bigg(q_n\ln q_n+\bar q_n\ln \bar q_n\bigg)\nonumber \\
&-\ln \int_{x_0}^{\frac{1}{2}} dx (q_n+\bar q_n) \int_{x_0}^{\frac{1}{2}} dx (q_n+\bar q_n) -\ln \int_{0}^{x_0-\delta} dx (q_n+\bar q_n) \int_{0}^{x_0-\delta} dx (q_n+\bar q_n) \ .
\end{align}
Clearly, it interpolates between the two special cases considered above. When $\delta=x_0$, it reduces to the totally asymmetric case,
while for $\delta=0$, it reduces to the symmetric case. The coefficient of the $\Lambda^-$ measures this asymmetry
\begin{align}
\frac{dS_n(x_0,\delta)}{d\ln \Lambda^-}=\int_{x_0-\delta}^{x_0}dx \bigg(q_n(x)+\bar q_n(x)\bigg) \ ,
\end{align}
which is always less or equal to $1$ but non-negative. It vanishes only for $\delta=0$.

\section{Naive bound for an $n$-parton state}~\label{App_generic}
Consider a generic wave function with maximally $n$-partons
\begin{align}
|\Phi\rangle=\sum_{i=1}^{n}\frac{1}{\sqrt{\Lambda^-}^i}\sum_{x_1,...x_i}\varphi_i(x_1,....x_i)|x_1,....x_i\rangle
\end{align}
After tracing over $A=[x_0,1-x_0]$, the reduced density matrix has the form
\begin{align}
\hat \rho_A=\sum_{i,j}\rho_{ij}|i\rangle \langle j| \ ,
\end{align}
with  $|i \rangle$  a generic state with $i$ particles.  It is important to observe,  that the diagonal terms
follow from tracing $|i\rangle \langle i|$, since partial tracing cannot change the difference in particle numbers. Therefore,
 the diagonal terms form  a reduced density matrix $\hat \rho_{A\rm dia}$, which contains more entropy compared to the full reduced density matrix, in general.
They could be used to derive a super-bound. Specifically, if we retain  only the diagonal terms, the reduced density matrix reads
\begin{align}
&\hat \rho_{A\rm dia}=\sum_{i=0}^{n-1}\frac{1}{(\Lambda^-)^{i-1}}|x_1,x_2,....x_i\rangle \langle x'_{1},x'_{2},...x'_{i}| \nonumber \\
&\times \sum_{j=i+1}^{n} \int_{y\in E^j_i(x,x')} dy \,\varphi_{j}^{\dagger}(x_1,...x_i;y_{i+1},...y_j)\varphi_{j}(x'_1,...x'_i;y_{i+1},..y_j) \ ,
\end{align}
with  $E^j_i(x,x')$ the region of the $j$-particle phase space which should be traced over, for a reduction to the
 $i$-body phase spaces. Performing another diagonal approximation, the entanglement  entropy  is bounded by
\begin{align}
S_E(x_0)\le \ln \Lambda^- \sum_{i=0}^{n-1} i p_i(x_0)+C \ .
\end{align}
Here, $p_i(x_0)$ is the sum of the cumulative probabilities
\begin{align}
p_i(x_0)=\sum_{j=i+1}^{n}\int_{x\in A_{i}^j(x_0)} dx_1,...x_j |\varphi_j(x_1,...x_j)|^2\ ,
\end{align}
where $A_{i}^{j}(x_0)$ is the part of the $j$-body phase space that after tracing, reduces to the
$i$-body state. While the sum over all the probabilities is less then $1$, the sum over the $i$-weighted probabilities is not a priori less than 1.

\section{Low-x analysis in 2D QCD}~\label{app_EXP}

The low-x analysis of the entanglement entropy in the single and asymmetric cut interval $A=[0,x_0\leq \frac 12]$,
can be carried exactly for $x_0\rightarrow 0$, in 2D QCD. More specifically, using (\ref{WFS}) allows to unwind
each contributions in  (\ref{SENTFULL})  as
\bea
&&\int_{0}^{x_0} dx \bigg[q_n(x)+\bar q_n(x)\bigg] =2C_n^2\frac{x_0^{2\beta+1}}{2\beta+1}+{\cal O}(x_0^2) \ , \nonumber\\
&&-\int_{0}^{x_0} dx\bigg[q_n(x)\ln q_n(x)+\bar q_n(x)\ln \bar q_n(x) \bigg]=4\beta C_n^2\frac{1+(2\beta+1)\ln \frac{1}{x_0}}{(2\beta+1)^2}x_0^{2\beta+1}+{\cal O}(x_0^2) \ ,  \nonumber\\
&&-\int_{x_0}^{\frac{1}{2}} dx \bigg[q_n(x)+\bar q_n(x)\bigg] \ln \int_{x_0}^{\frac{1}{2}} dx \bigg[q_n(x)+\bar q_n(x)\bigg] =2C_n^2\frac{x_0^{2\beta+1}}{2\beta+1}+{\cal O}(x_0^2) \ ,
\eea
and therefore, for $\beta>0$

\bea
\label{SNX0}
S_n(x_0)=2C_n^2\frac{x_0^{2\beta+1}}{2\beta+1}
\bigg(\ln (e\Lambda^-)+2\beta\frac{1+(2\beta+1)\ln \frac{1}{x_0}}{(2\beta+1)}+{\cal O}(x_0^2)\bigg) \ ,
\eea
which is the result quoted in the text.

\bibliography{ENT}

\begin{thebibliography}{32}%
\makeatletter
\providecommand \@ifxundefined [1]{%
 \@ifx{#1\undefined}
}%
\providecommand \@ifnum [1]{%
 \ifnum #1\expandafter \@firstoftwo
 \else \expandafter \@secondoftwo
 \fi
}%
\providecommand \@ifx [1]{%
 \ifx #1\expandafter \@firstoftwo
 \else \expandafter \@secondoftwo
 \fi
}%
\providecommand \natexlab [1]{#1}%
\providecommand \enquote  [1]{``#1''}%
\providecommand \bibnamefont  [1]{#1}%
\providecommand \bibfnamefont [1]{#1}%
\providecommand \citenamefont [1]{#1}%
\providecommand \href@noop [0]{\@secondoftwo}%
\providecommand \href [0]{\begingroup \@sanitize@url \@href}%
\providecommand \@href[1]{\@@startlink{#1}\@@href}%
\providecommand \@@href[1]{\endgroup#1\@@endlink}%
\providecommand \@sanitize@url [0]{\catcode `\\12\catcode `\$12\catcode
  `\&12\catcode `\#12\catcode `\^12\catcode `\_12\catcode `\%12\relax}%
\providecommand \@@startlink[1]{}%
\providecommand \@@endlink[0]{}%
\providecommand \url  [0]{\begingroup\@sanitize@url \@url }%
\providecommand \@url [1]{\endgroup\@href {#1}{\urlprefix }}%
\providecommand \urlprefix  [0]{URL }%
\providecommand \Eprint [0]{\href }%
\providecommand \doibase [0]{http://dx.doi.org/}%
\providecommand \selectlanguage [0]{\@gobble}%
\providecommand \bibinfo  [0]{\@secondoftwo}%
\providecommand \bibfield  [0]{\@secondoftwo}%
\providecommand \translation [1]{[#1]}%
\providecommand \BibitemOpen [0]{}%
\providecommand \bibitemStop [0]{}%
\providecommand \bibitemNoStop [0]{.\EOS\space}%
\providecommand \EOS [0]{\spacefactor3000\relax}%
\providecommand \BibitemShut  [1]{\csname bibitem#1\endcsname}%
\let\auto@bib@innerbib\@empty
\bibitem [{\citenamefont {Srednicki}(1993)}]{Srednicki:1993im}%
  \BibitemOpen
  \bibfield  {author} {\bibinfo {author} {\bibfnamefont {Mark}\ \bibnamefont
  {Srednicki}},\ }\bibfield  {title} {\enquote {\bibinfo {title} {{Entropy and
  area}},}\ }\href {\doibase 10.1103/PhysRevLett.71.666} {\bibfield  {journal}
  {\bibinfo  {journal} {Phys. Rev. Lett.}\ }\textbf {\bibinfo {volume} {71}},\
  \bibinfo {pages} {666--669} (\bibinfo {year} {1993})},\ \Eprint
  {http://arxiv.org/abs/hep-th/9303048} {arXiv:hep-th/9303048} \BibitemShut
  {NoStop}%
\bibitem [{\citenamefont {Calabrese}\ and\ \citenamefont
  {Cardy}(2004)}]{Calabrese:2004eu}%
  \BibitemOpen
  \bibfield  {author} {\bibinfo {author} {\bibfnamefont {Pasquale}\
  \bibnamefont {Calabrese}}\ and\ \bibinfo {author} {\bibfnamefont {John~L.}\
  \bibnamefont {Cardy}},\ }\bibfield  {title} {\enquote {\bibinfo {title}
  {{Entanglement entropy and quantum field theory}},}\ }\href {\doibase
  10.1088/1742-5468/2004/06/P06002} {\bibfield  {journal} {\bibinfo  {journal}
  {J. Stat. Mech.}\ }\textbf {\bibinfo {volume} {0406}},\ \bibinfo {pages}
  {P06002} (\bibinfo {year} {2004})},\ \Eprint
  {http://arxiv.org/abs/hep-th/0405152} {arXiv:hep-th/0405152} \BibitemShut
  {NoStop}%
\bibitem [{\citenamefont {Casini}\ \emph {et~al.}(2005)\citenamefont {Casini},
  \citenamefont {Fosco},\ and\ \citenamefont {Huerta}}]{Casini:2005rm}%
  \BibitemOpen
  \bibfield  {author} {\bibinfo {author} {\bibfnamefont {H.}~\bibnamefont
  {Casini}}, \bibinfo {author} {\bibfnamefont {C.~D.}\ \bibnamefont {Fosco}}, \
  and\ \bibinfo {author} {\bibfnamefont {M.}~\bibnamefont {Huerta}},\
  }\bibfield  {title} {\enquote {\bibinfo {title} {{Entanglement and alpha
  entropies for a massive Dirac field in two dimensions}},}\ }\href {\doibase
  10.1088/1742-5468/2005/07/P07007} {\bibfield  {journal} {\bibinfo  {journal}
  {J. Stat. Mech.}\ }\textbf {\bibinfo {volume} {0507}},\ \bibinfo {pages}
  {P07007} (\bibinfo {year} {2005})},\ \Eprint
  {http://arxiv.org/abs/cond-mat/0505563} {arXiv:cond-mat/0505563} \BibitemShut
  {NoStop}%
\bibitem [{\citenamefont {Hastings}(2007)}]{Hastings:2007iok}%
  \BibitemOpen
  \bibfield  {author} {\bibinfo {author} {\bibfnamefont {M.~B.}\ \bibnamefont
  {Hastings}},\ }\bibfield  {title} {\enquote {\bibinfo {title} {{An area law
  for one-dimensional quantum systems}},}\ }\href {\doibase
  10.1088/1742-5468/2007/08/P08024} {\bibfield  {journal} {\bibinfo  {journal}
  {J. Stat. Mech.}\ }\textbf {\bibinfo {volume} {0708}},\ \bibinfo {pages}
  {P08024} (\bibinfo {year} {2007})},\ \Eprint {http://arxiv.org/abs/0705.2024}
  {arXiv:0705.2024 [quant-ph]} \BibitemShut {NoStop}%
\bibitem [{\citenamefont {Calabrese}\ and\ \citenamefont
  {Cardy}(2009)}]{Calabrese:2009qy}%
  \BibitemOpen
  \bibfield  {author} {\bibinfo {author} {\bibfnamefont {Pasquale}\
  \bibnamefont {Calabrese}}\ and\ \bibinfo {author} {\bibfnamefont {John}\
  \bibnamefont {Cardy}},\ }\bibfield  {title} {\enquote {\bibinfo {title}
  {{Entanglement entropy and conformal field theory}},}\ }\href {\doibase
  10.1088/1751-8113/42/50/504005} {\bibfield  {journal} {\bibinfo  {journal}
  {J. Phys. A}\ }\textbf {\bibinfo {volume} {42}},\ \bibinfo {pages} {504005}
  (\bibinfo {year} {2009})},\ \Eprint {http://arxiv.org/abs/0905.4013}
  {arXiv:0905.4013 [cond-mat.stat-mech]} \BibitemShut {NoStop}%
\bibitem [{\citenamefont {Bremermann}(1967)}]{le1967proceedings}%
  \BibitemOpen
  \bibfield  {author} {\bibinfo {author} {\bibfnamefont {H.J.}\ \bibnamefont
  {Bremermann}},\ }\href@noop {} {\emph {\bibinfo {title} {in Proceedings of
  the Fifth Berkeley Symposium on Mathematical Statistics and Probability, Eds.
  Le Cam, Lucien Marie and Neyman, Jerzy}}},\ Vol.~\bibinfo {volume} {3}\
  (\bibinfo  {publisher} {Univ of California Press},\ \bibinfo {year}
  {1967})\BibitemShut {NoStop}%
\bibitem [{\citenamefont {Bekenstein}(1981)}]{Bekenstein:1981zz}%
  \BibitemOpen
  \bibfield  {author} {\bibinfo {author} {\bibfnamefont {Jacob~D.}\
  \bibnamefont {Bekenstein}},\ }\bibfield  {title} {\enquote {\bibinfo {title}
  {{Energy Cost of Information Transfer}},}\ }\href {\doibase
  10.1103/PhysRevLett.46.623} {\bibfield  {journal} {\bibinfo  {journal} {Phys.
  Rev. Lett.}\ }\textbf {\bibinfo {volume} {46}},\ \bibinfo {pages} {623--626}
  (\bibinfo {year} {1981})}\BibitemShut {NoStop}%
\bibitem [{\citenamefont {Feynman}(1969)}]{Feynman:1969wa}%
  \BibitemOpen
  \bibfield  {author} {\bibinfo {author} {\bibfnamefont {R.~P.}\ \bibnamefont
  {Feynman}},\ }\bibfield  {title} {\enquote {\bibinfo {title} {{The behavior
  of hadron collisions at extreme energies}},}\ }\href@noop {} {\bibfield
  {journal} {\bibinfo  {journal} {Conf. Proc. C}\ }\textbf {\bibinfo {volume}
  {690905}},\ \bibinfo {pages} {237--258} (\bibinfo {year} {1969})}\BibitemShut
  {NoStop}%
\bibitem [{\citenamefont {Susskind}(1993)}]{Susskind:1993ki}%
  \BibitemOpen
  \bibfield  {author} {\bibinfo {author} {\bibfnamefont {Leonard}\ \bibnamefont
  {Susskind}},\ }\bibfield  {title} {\enquote {\bibinfo {title} {{String theory
  and the principles of black hole complementarity}},}\ }\href {\doibase
  10.1103/PhysRevLett.71.2367} {\bibfield  {journal} {\bibinfo  {journal}
  {Phys. Rev. Lett.}\ }\textbf {\bibinfo {volume} {71}},\ \bibinfo {pages}
  {2367--2368} (\bibinfo {year} {1993})},\ \Eprint
  {http://arxiv.org/abs/hep-th/9307168} {arXiv:hep-th/9307168} \BibitemShut
  {NoStop}%
\bibitem [{\citenamefont {Susskind}(1994)}]{Susskind:1993aa}%
  \BibitemOpen
  \bibfield  {author} {\bibinfo {author} {\bibfnamefont {Leonard}\ \bibnamefont
  {Susskind}},\ }\bibfield  {title} {\enquote {\bibinfo {title} {{Strings,
  black holes and Lorentz contraction}},}\ }\href {\doibase
  10.1103/PhysRevD.49.6606} {\bibfield  {journal} {\bibinfo  {journal} {Phys.
  Rev. D}\ }\textbf {\bibinfo {volume} {49}},\ \bibinfo {pages} {6606--6611}
  (\bibinfo {year} {1994})},\ \Eprint {http://arxiv.org/abs/hep-th/9308139}
  {arXiv:hep-th/9308139} \BibitemShut {NoStop}%
\bibitem [{\citenamefont {Thorn}(1995)}]{Thorn:1994sw}%
  \BibitemOpen
  \bibfield  {author} {\bibinfo {author} {\bibfnamefont {Charles~B.}\
  \bibnamefont {Thorn}},\ }\bibfield  {title} {\enquote {\bibinfo {title}
  {{Calculating the rest tension for a polymer of string bits}},}\ }\href
  {\doibase 10.1103/PhysRevD.51.647} {\bibfield  {journal} {\bibinfo  {journal}
  {Phys. Rev. D}\ }\textbf {\bibinfo {volume} {51}},\ \bibinfo {pages}
  {647--664} (\bibinfo {year} {1995})},\ \Eprint
  {http://arxiv.org/abs/hep-th/9407169} {arXiv:hep-th/9407169} \BibitemShut
  {NoStop}%
\bibitem [{\citenamefont {Stoffers}\ and\ \citenamefont
  {Zahed}(2013{\natexlab{a}})}]{Stoffers:2012mn}%
  \BibitemOpen
  \bibfield  {author} {\bibinfo {author} {\bibfnamefont {Alexander}\
  \bibnamefont {Stoffers}}\ and\ \bibinfo {author} {\bibfnamefont {Ismail}\
  \bibnamefont {Zahed}},\ }\bibfield  {title} {\enquote {\bibinfo {title}
  {{Holographic Pomeron and Entropy}},}\ }\href {\doibase
  10.1103/PhysRevD.88.025038} {\bibfield  {journal} {\bibinfo  {journal} {Phys.
  Rev. D}\ }\textbf {\bibinfo {volume} {88}},\ \bibinfo {pages} {025038}
  (\bibinfo {year} {2013}{\natexlab{a}})},\ \Eprint
  {http://arxiv.org/abs/1211.3077} {arXiv:1211.3077 [nucl-th]} \BibitemShut
  {NoStop}%
\bibitem [{\citenamefont {Shuryak}\ and\ \citenamefont
  {Zahed}(2018)}]{Shuryak:2017phz}%
  \BibitemOpen
  \bibfield  {author} {\bibinfo {author} {\bibfnamefont {Edward}\ \bibnamefont
  {Shuryak}}\ and\ \bibinfo {author} {\bibfnamefont {Ismail}\ \bibnamefont
  {Zahed}},\ }\bibfield  {title} {\enquote {\bibinfo {title} {{Regimes of the
  Pomeron and its Intrinsic Entropy}},}\ }\href {\doibase
  10.1016/j.aop.2018.06.008} {\bibfield  {journal} {\bibinfo  {journal} {Annals
  Phys.}\ }\textbf {\bibinfo {volume} {396}},\ \bibinfo {pages} {1--17}
  (\bibinfo {year} {2018})},\ \Eprint {http://arxiv.org/abs/1707.01885}
  {arXiv:1707.01885 [hep-ph]} \BibitemShut {NoStop}%
\bibitem [{\citenamefont {Liu}\ and\ \citenamefont
  {Zahed}(2019)}]{Liu:2018gae}%
  \BibitemOpen
  \bibfield  {author} {\bibinfo {author} {\bibfnamefont {Yizhuang}\
  \bibnamefont {Liu}}\ and\ \bibinfo {author} {\bibfnamefont {Ismail}\
  \bibnamefont {Zahed}},\ }\bibfield  {title} {\enquote {\bibinfo {title}
  {{Entanglement in Regge scattering using the AdS/CFT correspondence}},}\
  }\href {\doibase 10.1103/PhysRevD.100.046005} {\bibfield  {journal} {\bibinfo
   {journal} {Phys. Rev. D}\ }\textbf {\bibinfo {volume} {100}},\ \bibinfo
  {pages} {046005} (\bibinfo {year} {2019})},\ \Eprint
  {http://arxiv.org/abs/1803.09157} {arXiv:1803.09157 [hep-ph]} \BibitemShut
  {NoStop}%
\bibitem [{\citenamefont {Kharzeev}\ and\ \citenamefont
  {Levin}(2017)}]{Kharzeev:2017qzs}%
  \BibitemOpen
  \bibfield  {author} {\bibinfo {author} {\bibfnamefont {Dmitri~E.}\
  \bibnamefont {Kharzeev}}\ and\ \bibinfo {author} {\bibfnamefont {Eugene~M.}\
  \bibnamefont {Levin}},\ }\bibfield  {title} {\enquote {\bibinfo {title}
  {{Deep inelastic scattering as a probe of entanglement}},}\ }\href {\doibase
  10.1103/PhysRevD.95.114008} {\bibfield  {journal} {\bibinfo  {journal} {Phys.
  Rev. D}\ }\textbf {\bibinfo {volume} {95}},\ \bibinfo {pages} {114008}
  (\bibinfo {year} {2017})},\ \Eprint {http://arxiv.org/abs/1702.03489}
  {arXiv:1702.03489 [hep-ph]} \BibitemShut {NoStop}%
\bibitem [{\citenamefont {Armesto}\ \emph {et~al.}(2019)\citenamefont
  {Armesto}, \citenamefont {Dominguez}, \citenamefont {Kovner}, \citenamefont
  {Lublinsky},\ and\ \citenamefont {Skokov}}]{Armesto:2019mna}%
  \BibitemOpen
  \bibfield  {author} {\bibinfo {author} {\bibfnamefont {Nestor}\ \bibnamefont
  {Armesto}}, \bibinfo {author} {\bibfnamefont {Fabio}\ \bibnamefont
  {Dominguez}}, \bibinfo {author} {\bibfnamefont {Alex}\ \bibnamefont
  {Kovner}}, \bibinfo {author} {\bibfnamefont {Michael}\ \bibnamefont
  {Lublinsky}}, \ and\ \bibinfo {author} {\bibfnamefont {Vladimir}\
  \bibnamefont {Skokov}},\ }\bibfield  {title} {\enquote {\bibinfo {title}
  {{The Color Glass Condensate density matrix: Lindblad evolution, entanglement
  entropy and Wigner functional}},}\ }\href {\doibase 10.1007/JHEP05(2019)025}
  {\bibfield  {journal} {\bibinfo  {journal} {JHEP}\ }\textbf {\bibinfo
  {volume} {05}},\ \bibinfo {pages} {025} (\bibinfo {year} {2019})},\ \Eprint
  {http://arxiv.org/abs/1901.08080} {arXiv:1901.08080 [hep-ph]} \BibitemShut
  {NoStop}%
\bibitem [{\citenamefont {Dvali}\ and\ \citenamefont
  {Venugopalan}(2021)}]{Dvali:2021ooc}%
  \BibitemOpen
  \bibfield  {author} {\bibinfo {author} {\bibfnamefont {Gia}\ \bibnamefont
  {Dvali}}\ and\ \bibinfo {author} {\bibfnamefont {Raju}\ \bibnamefont
  {Venugopalan}},\ }\bibfield  {title} {\enquote {\bibinfo {title}
  {{Classicalization and unitarization of wee partons in QCD and Gravity: The
  CGC-Black Hole correspondence}},}\ }\href@noop {} {\  (\bibinfo {year}
  {2021})},\ \Eprint {http://arxiv.org/abs/2106.11989} {arXiv:2106.11989
  [hep-th]} \BibitemShut {NoStop}%
\bibitem [{\citenamefont {Qian}\ and\ \citenamefont
  {Zahed}(2015{\natexlab{a}})}]{Qian:2015boa}%
  \BibitemOpen
  \bibfield  {author} {\bibinfo {author} {\bibfnamefont {Yachao}\ \bibnamefont
  {Qian}}\ and\ \bibinfo {author} {\bibfnamefont {Ismail}\ \bibnamefont
  {Zahed}},\ }\bibfield  {title} {\enquote {\bibinfo {title} {{Stretched string
  with self-interaction at the Hagedorn point: Spatial sizes and black
  holes}},}\ }\href {\doibase 10.1103/PhysRevD.92.105001} {\bibfield  {journal}
  {\bibinfo  {journal} {Phys. Rev. D}\ }\textbf {\bibinfo {volume} {92}},\
  \bibinfo {pages} {105001} (\bibinfo {year} {2015}{\natexlab{a}})},\ \Eprint
  {http://arxiv.org/abs/1508.03760} {arXiv:1508.03760 [hep-ph]} \BibitemShut
  {NoStop}%
\bibitem [{\citenamefont {Karliner}\ \emph {et~al.}(1988)\citenamefont
  {Karliner}, \citenamefont {Klebanov},\ and\ \citenamefont
  {Susskind}}]{Karliner:1988hd}%
  \BibitemOpen
  \bibfield  {author} {\bibinfo {author} {\bibfnamefont {Marek}\ \bibnamefont
  {Karliner}}, \bibinfo {author} {\bibfnamefont {Igor~R.}\ \bibnamefont
  {Klebanov}}, \ and\ \bibinfo {author} {\bibfnamefont {Leonard}\ \bibnamefont
  {Susskind}},\ }\bibfield  {title} {\enquote {\bibinfo {title} {{Size and
  Shape of Strings}},}\ }\href {\doibase 10.1142/S0217751X88000837} {\bibfield
  {journal} {\bibinfo  {journal} {Int. J. Mod. Phys. A}\ }\textbf {\bibinfo
  {volume} {3}},\ \bibinfo {pages} {1981} (\bibinfo {year} {1988})}\BibitemShut
  {NoStop}%
\bibitem [{\citenamefont {Qian}\ and\ \citenamefont
  {Zahed}(2015{\natexlab{b}})}]{Qian:2014rda}%
  \BibitemOpen
  \bibfield  {author} {\bibinfo {author} {\bibfnamefont {Yachao}\ \bibnamefont
  {Qian}}\ and\ \bibinfo {author} {\bibfnamefont {Ismail}\ \bibnamefont
  {Zahed}},\ }\bibfield  {title} {\enquote {\bibinfo {title} {{Stretched String
  with Self-Interaction at High Resolution: Spatial Sizes and Saturation}},}\
  }\href {\doibase 10.1103/PhysRevD.91.125032} {\bibfield  {journal} {\bibinfo
  {journal} {Phys. Rev. D}\ }\textbf {\bibinfo {volume} {91}},\ \bibinfo
  {pages} {125032} (\bibinfo {year} {2015}{\natexlab{b}})},\ \Eprint
  {http://arxiv.org/abs/1411.3653} {arXiv:1411.3653 [hep-ph]} \BibitemShut
  {NoStop}%
\bibitem [{\citenamefont {'t~Hooft}(1974)}]{tHooft:1974pnl}%
  \BibitemOpen
  \bibfield  {author} {\bibinfo {author} {\bibfnamefont {Gerard}\ \bibnamefont
  {'t~Hooft}},\ }\bibfield  {title} {\enquote {\bibinfo {title} {{A
  Two-Dimensional Model for Mesons}},}\ }\href {\doibase
  10.1016/0550-3213(74)90088-1} {\bibfield  {journal} {\bibinfo  {journal}
  {Nucl. Phys. B}\ }\textbf {\bibinfo {volume} {75}},\ \bibinfo {pages}
  {461--470} (\bibinfo {year} {1974})}\BibitemShut {NoStop}%
\bibitem [{\citenamefont {Bars}(1976)}]{Bars:1976nk}%
  \BibitemOpen
  \bibfield  {author} {\bibinfo {author} {\bibfnamefont {I.}~\bibnamefont
  {Bars}},\ }\bibfield  {title} {\enquote {\bibinfo {title} {{A Quantum String
  Theory of Hadrons and Its Relation to Quantum Chromodynamics in
  Two-Dimensions}},}\ }\href {\doibase 10.1016/0550-3213(76)90327-8} {\bibfield
   {journal} {\bibinfo  {journal} {Nucl. Phys. B}\ }\textbf {\bibinfo {volume}
  {111}},\ \bibinfo {pages} {413--440} (\bibinfo {year} {1976})}\BibitemShut
  {NoStop}%
\bibitem [{\citenamefont {Latora}\ \emph {et~al.}(2000)\citenamefont {Latora},
  \citenamefont {Rapisarda}, \citenamefont {Tsallis},\ and\ \citenamefont
  {Baranger}}]{Latora:1999vk}%
  \BibitemOpen
  \bibfield  {author} {\bibinfo {author} {\bibfnamefont {Vito}\ \bibnamefont
  {Latora}}, \bibinfo {author} {\bibfnamefont {Andrea}\ \bibnamefont
  {Rapisarda}}, \bibinfo {author} {\bibfnamefont {Constantino}\ \bibnamefont
  {Tsallis}}, \ and\ \bibinfo {author} {\bibfnamefont {Michel}\ \bibnamefont
  {Baranger}},\ }\bibfield  {title} {\enquote {\bibinfo {title}
  {{Generalization to nonextensive systems of the rate of entropy increase: The
  Case of the logistic map}},}\ }\href {\doibase 10.1016/S0375-9601(00)00484-9}
  {\bibfield  {journal} {\bibinfo  {journal} {Phys. Lett. A}\ }\textbf
  {\bibinfo {volume} {273}},\ \bibinfo {pages} {97} (\bibinfo {year} {2000})},\
  \Eprint {http://arxiv.org/abs/cond-mat/9907412} {arXiv:cond-mat/9907412}
  \BibitemShut {NoStop}%
\bibitem [{\citenamefont {Pauli}\ and\ \citenamefont
  {Brodsky}(1985)}]{Pauli:1985ps}%
  \BibitemOpen
  \bibfield  {author} {\bibinfo {author} {\bibfnamefont {Hans~Christian}\
  \bibnamefont {Pauli}}\ and\ \bibinfo {author} {\bibfnamefont {Stanley~J.}\
  \bibnamefont {Brodsky}},\ }\bibfield  {title} {\enquote {\bibinfo {title}
  {{Discretized Light Cone Quantization: Solution to a Field Theory in One
  Space One Time Dimensions}},}\ }\href {\doibase 10.1103/PhysRevD.32.2001}
  {\bibfield  {journal} {\bibinfo  {journal} {Phys. Rev. D}\ }\textbf {\bibinfo
  {volume} {32}},\ \bibinfo {pages} {2001} (\bibinfo {year}
  {1985})}\BibitemShut {NoStop}%
\bibitem [{\citenamefont {Eller}\ \emph {et~al.}(1987)\citenamefont {Eller},
  \citenamefont {Pauli},\ and\ \citenamefont {Brodsky}}]{Eller:1986nt}%
  \BibitemOpen
  \bibfield  {author} {\bibinfo {author} {\bibfnamefont {Thomas}\ \bibnamefont
  {Eller}}, \bibinfo {author} {\bibfnamefont {Hans~Christian}\ \bibnamefont
  {Pauli}}, \ and\ \bibinfo {author} {\bibfnamefont {Stanley~J.}\ \bibnamefont
  {Brodsky}},\ }\bibfield  {title} {\enquote {\bibinfo {title} {{Discretized
  Light Cone Quantization: The Massless and the Massive Schwinger Model}},}\
  }\href {\doibase 10.1103/PhysRevD.35.1493} {\bibfield  {journal} {\bibinfo
  {journal} {Phys. Rev. D}\ }\textbf {\bibinfo {volume} {35}},\ \bibinfo
  {pages} {1493} (\bibinfo {year} {1987})}\BibitemShut {NoStop}%
\bibitem [{\citenamefont {Eller}\ and\ \citenamefont
  {Pauli}(1989)}]{Eller:1989ek}%
  \BibitemOpen
  \bibfield  {author} {\bibinfo {author} {\bibfnamefont {T.}~\bibnamefont
  {Eller}}\ and\ \bibinfo {author} {\bibfnamefont {H.~C.}\ \bibnamefont
  {Pauli}},\ }\bibfield  {title} {\enquote {\bibinfo {title} {{Quantizing {QED}
  in Two-dimensions on the Light Cone}},}\ }\href {\doibase 10.1007/BF01565128}
  {\bibfield  {journal} {\bibinfo  {journal} {Z. Phys. C}\ }\textbf {\bibinfo
  {volume} {42}},\ \bibinfo {pages} {59--67} (\bibinfo {year}
  {1989})}\BibitemShut {NoStop}%
\bibitem [{\citenamefont {Kharzeev}(2021)}]{Kharzeev:2021nzh}%
  \BibitemOpen
  \bibfield  {author} {\bibinfo {author} {\bibfnamefont {Dmitri~E.}\
  \bibnamefont {Kharzeev}},\ }\bibfield  {title} {\enquote {\bibinfo {title}
  {{Quantum information approach to high energy interactions}},}\ }\href
  {\doibase 10.1098/rsta.2021.0063} {\bibfield  {journal} {\bibinfo  {journal}
  {Phil. Trans. A. Math. Phys. Eng. Sci.}\ }\textbf {\bibinfo {volume} {380}},\
  \bibinfo {pages} {20210063} (\bibinfo {year} {2021})},\ \Eprint
  {http://arxiv.org/abs/2108.08792} {arXiv:2108.08792 [hep-ph]} \BibitemShut
  {NoStop}%
\bibitem [{\citenamefont {Brower}\ \emph {et~al.}(1979)\citenamefont {Brower},
  \citenamefont {Spence},\ and\ \citenamefont {Weis}}]{Brower:1978wm}%
  \BibitemOpen
  \bibfield  {author} {\bibinfo {author} {\bibfnamefont {R.~C.}\ \bibnamefont
  {Brower}}, \bibinfo {author} {\bibfnamefont {W.~L.}\ \bibnamefont {Spence}},
  \ and\ \bibinfo {author} {\bibfnamefont {J.~H.}\ \bibnamefont {Weis}},\
  }\bibfield  {title} {\enquote {\bibinfo {title} {{Bound States and Asymptotic
  Limits for {QCD} in Two-dimensions}},}\ }\href {\doibase
  10.1103/PhysRevD.19.3024} {\bibfield  {journal} {\bibinfo  {journal} {Phys.
  Rev. D}\ }\textbf {\bibinfo {volume} {19}},\ \bibinfo {pages} {3024}
  (\bibinfo {year} {1979})}\BibitemShut {NoStop}%
\bibitem [{\citenamefont {Katz}\ and\ \citenamefont
  {Okui}(2009)}]{Katz:2007br}%
  \BibitemOpen
  \bibfield  {author} {\bibinfo {author} {\bibfnamefont {Emanuel}\ \bibnamefont
  {Katz}}\ and\ \bibinfo {author} {\bibfnamefont {Takemichi}\ \bibnamefont
  {Okui}},\ }\bibfield  {title} {\enquote {\bibinfo {title} {{The 't Hooft
  model as a hologram}},}\ }\href {\doibase 10.1088/1126-6708/2009/01/013}
  {\bibfield  {journal} {\bibinfo  {journal} {JHEP}\ }\textbf {\bibinfo
  {volume} {01}},\ \bibinfo {pages} {013} (\bibinfo {year} {2009})},\ \Eprint
  {http://arxiv.org/abs/0710.3402} {arXiv:0710.3402 [hep-th]} \BibitemShut
  {NoStop}%
\bibitem [{\citenamefont {Kock}\ \emph {et~al.}(2020)\citenamefont {Kock},
  \citenamefont {Liu},\ and\ \citenamefont {Zahed}}]{Kock:2020frx}%
  \BibitemOpen
  \bibfield  {author} {\bibinfo {author} {\bibfnamefont {Arthur}\ \bibnamefont
  {Kock}}, \bibinfo {author} {\bibfnamefont {Yizhuang}\ \bibnamefont {Liu}}, \
  and\ \bibinfo {author} {\bibfnamefont {Ismail}\ \bibnamefont {Zahed}},\
  }\bibfield  {title} {\enquote {\bibinfo {title} {{Pion and kaon parton
  distributions in the QCD instanton vacuum}},}\ }\href {\doibase
  10.1103/PhysRevD.102.014039} {\bibfield  {journal} {\bibinfo  {journal}
  {Phys. Rev. D}\ }\textbf {\bibinfo {volume} {102}},\ \bibinfo {pages}
  {014039} (\bibinfo {year} {2020})},\ \Eprint
  {http://arxiv.org/abs/2004.01595} {arXiv:2004.01595 [hep-ph]} \BibitemShut
  {NoStop}%
\bibitem [{\citenamefont {Bergman}\ and\ \citenamefont
  {Thorn}(1997)}]{Bergman:1997ki}%
  \BibitemOpen
  \bibfield  {author} {\bibinfo {author} {\bibfnamefont {Oren}\ \bibnamefont
  {Bergman}}\ and\ \bibinfo {author} {\bibfnamefont {Charles~B.}\ \bibnamefont
  {Thorn}},\ }\bibfield  {title} {\enquote {\bibinfo {title} {{The Size of a
  polymer of string bits: A Numerical investigation}},}\ }\href {\doibase
  10.1016/S0550-3213(97)00475-6} {\bibfield  {journal} {\bibinfo  {journal}
  {Nucl. Phys. B}\ }\textbf {\bibinfo {volume} {502}},\ \bibinfo {pages}
  {309--324} (\bibinfo {year} {1997})},\ \Eprint
  {http://arxiv.org/abs/hep-th/9702068} {arXiv:hep-th/9702068} \BibitemShut
  {NoStop}%
\bibitem [{\citenamefont {Stoffers}\ and\ \citenamefont
  {Zahed}(2013{\natexlab{b}})}]{Stoffers:2012zw}%
  \BibitemOpen
  \bibfield  {author} {\bibinfo {author} {\bibfnamefont {Alexander}\
  \bibnamefont {Stoffers}}\ and\ \bibinfo {author} {\bibfnamefont {Ismail}\
  \bibnamefont {Zahed}},\ }\bibfield  {title} {\enquote {\bibinfo {title}
  {{Holographic Pomeron: Saturation and DIS}},}\ }\href {\doibase
  10.1103/PhysRevD.87.075023} {\bibfield  {journal} {\bibinfo  {journal} {Phys.
  Rev. D}\ }\textbf {\bibinfo {volume} {87}},\ \bibinfo {pages} {075023}
  (\bibinfo {year} {2013}{\natexlab{b}})},\ \Eprint
  {http://arxiv.org/abs/1205.3223} {arXiv:1205.3223 [hep-ph]} \BibitemShut
  {NoStop}%
\end{thebibliography}%

\end{document}